\documentclass[10pt]{IEEEtran}
\usepackage{amsmath}
\usepackage[T1]{fontenc}
\usepackage{algorithm}
\usepackage{alphalph}
\usepackage{algorithmic}
\usepackage{listings}
\usepackage{amsmath}
\usepackage{graphicx}
\usepackage{color}
\usepackage[cmintegrals]{newtxmath}
\usepackage[caption=false,font=footnotesize]{subfig}
\ifCLASSOPTIONcompsoc
\usepackage[caption=false,font=normalsize,labelfont=sf,textfont=sf]{subfig}
\else
\usepackage[caption=false,font=footnotesize]{subfig}
\fi
\usepackage[colorlinks,citecolor=red,urlcolor=blue,bookmarks=false,hypertexnames=true]{hyperref} 
\usepackage{caption}
\usepackage{lipsum}
\usepackage{nameref}
\usepackage{varioref}
\usepackage{hyperref}
\usepackage{setspace}

\begin{document}
\title{A Survey on Deep-Learning based Techniques for Modeling and Estimation of MassiveMIMO Channels}
\author{Makan~Zamanipour,~\IEEEmembership{Member,~IEEE}
\thanks{makan.zamanipour.2015@ieee.org; ORCID: 0000-0003-1606-9347; Researcher-ID: P-6298-2019. }

}

\maketitle
\markboth{IEEE, VOL. XX, NO. XX, X 2019}%
{Shell \MakeLowercase{\textit{et al.}}: Bare Demo of IEEEtran.cls for Computer Society Journals}
\begin{abstract}
\textit{Why does the literature consider the channel-state-information (CSI) as a 2/3-D image? What are the pros-and-cons of this consideration for accuracy-complexity trade-off?} Next generations of wireless communications require innumerable disciplines according to which a low-latency, low-traffic, high-throughput, high spectral-efficiency and low energy-consumption are guaranteed. Towards this end, the principle of massive multi-input multi-output (MaMIMO) is emerging which is conveniently deployed for millimeter wave (mmWave) bands. However, practical and realistic MaMIMO transceivers suffer from a huge range of challenging bottlenecks in design the majority of which belong to the issue of channel-estimation. Channel modeling and prediction in MaMIMO particularly suffer from computational complexity due to a high number of antenna sets and supported users. This complexity lies dominantly upon the feedback-overhead which even degrades the pilot-data trade-off in the uplink (UL)/downlink (DL) design. This comprehensive survey studies the novel deep-learning (DLg) driven techniques recently proposed in the literature which tackle the challenges discussed-above - which is for the first time. In addition, we consequently propose 7 open trends e.g. in the context of the lack of Q-learning in MaMIMO detection - for which we talk about a possible solution to the saddle-point in the 2-D pilot-data axis for a \textit{Stackelberg game} based scenario.
\end{abstract}


\maketitle

\IEEEdisplaynontitleabstractindextext
\IEEEpeerreviewmaketitle

\section{Introduction}
The fifth generation (5G) deployment is brilliant in various areas in relation to multi-standard and multi-functional wireless communications. The restricted design-budget and hardware resources cause a series of implementation bottlenecks at the user-equipment (UE) terminal. In relation to 5G standards, the principle of massive multi-input multi-output (MaMIMO) is emerging. MaMIMO dramatically enhances the spectral efficiency of the wireless network (Net) in addition to a guarantee for lower-latency and a higher energy efficiency.

The time-varying channel relating to each transmit-receive antenna pair is interpreted by the channel state-information (CSI) during the bandwidth (BW) of the system. In an orthogonal-frequency-division-multiplexing (OFDM)-based system, the CSI is an $N_t \times N_r \times T \times B$ \textit{tensor} \cite{R1R1} w.r.t. the channel matrix $\mathbf{H}$ of size $N_t \times N_r$. The values $N_t$, $N_r$, $B$ and $T$ stand respectively for the number of the transmit and receive antenna arrays, the numbers of subcarriers and OFDM symbols. The CSI is refreshed at a rate relying upon the correlation-time of the channel. For example, when CSI is low/high, the physical-layer needs to employ a low/high-order modulation scheme and a low/high coding rate for ensuring the bit-error-rate. Indeed, the CSI is a function of the attenuation of a radio signal over the air, that is, the impact of fading, path-loss, scattering, shadowing, atmosphere factors (rain, daytime based dynamicity/staticity obstacles, water vapor, molecules of oxygen, and other gaseous atmospheric components related to the time domain, air density, air humidity) etc \cite{R2R2}. After precisely considering all related factors, some certain patterns for CSI are observable. In particular, totally different frequency bands have totally different CSI even at the same location and at the same time-zone. Unfortunately, traditional equalisers fail for MaMIMO in millimeter waves (mmWaves). In fact, a simple linear channel-estimator\footnote{Such as zero-forcing.} cannot guarantee an acceptable minimum-mean-square-error (MMSE) over strong spatially correlated channels\footnote{A high number of antenna arrays in a high frequency carrier, succinctly speaking, makes the chipset less in size, leading to experience a correlated MaMIMO transceiver.}. Moreover, by the classical MMSE estimator with a high complexity only a suboptimal estimations can be guaranteed. Therefore, novel algorithms which can efficiently guarantee the complexity-accuracy trade-off are still required as favourable as possible. 

\begin{figure}[t]
\centering
{\includegraphics[trim={{42mm} {99 mm} {56 mm} {35mm}},clip,scale=0.4]{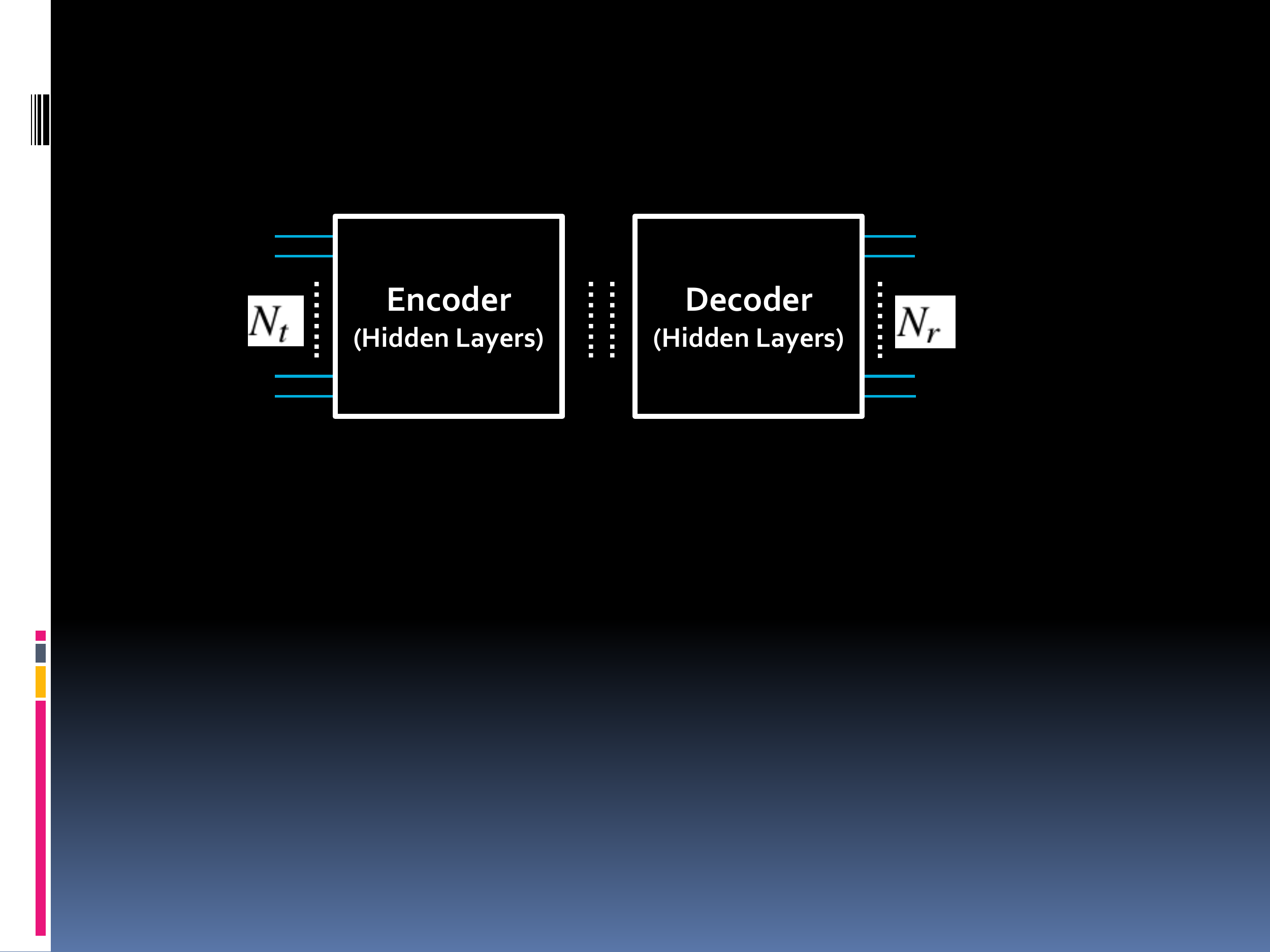}} 
\caption{A DNN for MaMIMO.}  \label{F1}
\label{fig:EcUND} 
\end{figure} 

There seems to be an enormous number of substantial tradeoffs among system parameters in the context of channel estimation, transmit-power allocation etc. These issues can be conveniently optimised through virtualising of tools according to deep-learning (DLg) based techniques. The basic architecture of a DLg mechanism is a multiple hidden-layer one. Indeed, any \textit{Borel} measurable function according to the \textit{universal approximation theorem}, can be \textit{inferred} by a DLg network (Net). DLg procedures can solve complex non-convexities according to which the model is considered as a black box. For a DLg architecture, activation functions optimise multiple layers of the Net resulting in the favourable and accurate maps. The logic behind of a DLg framework in MaMIMO also stems principally from the following fact: \textit{The channel matrix of beam-space for a mmWave MaMIMO transceiver can be taken into account in terms of a 2-D image.} 

\begin{figure}[t]
\centering
{\includegraphics[trim={{10mm} {0 mm} {1 mm} {10mm}},clip,scale=0.3]{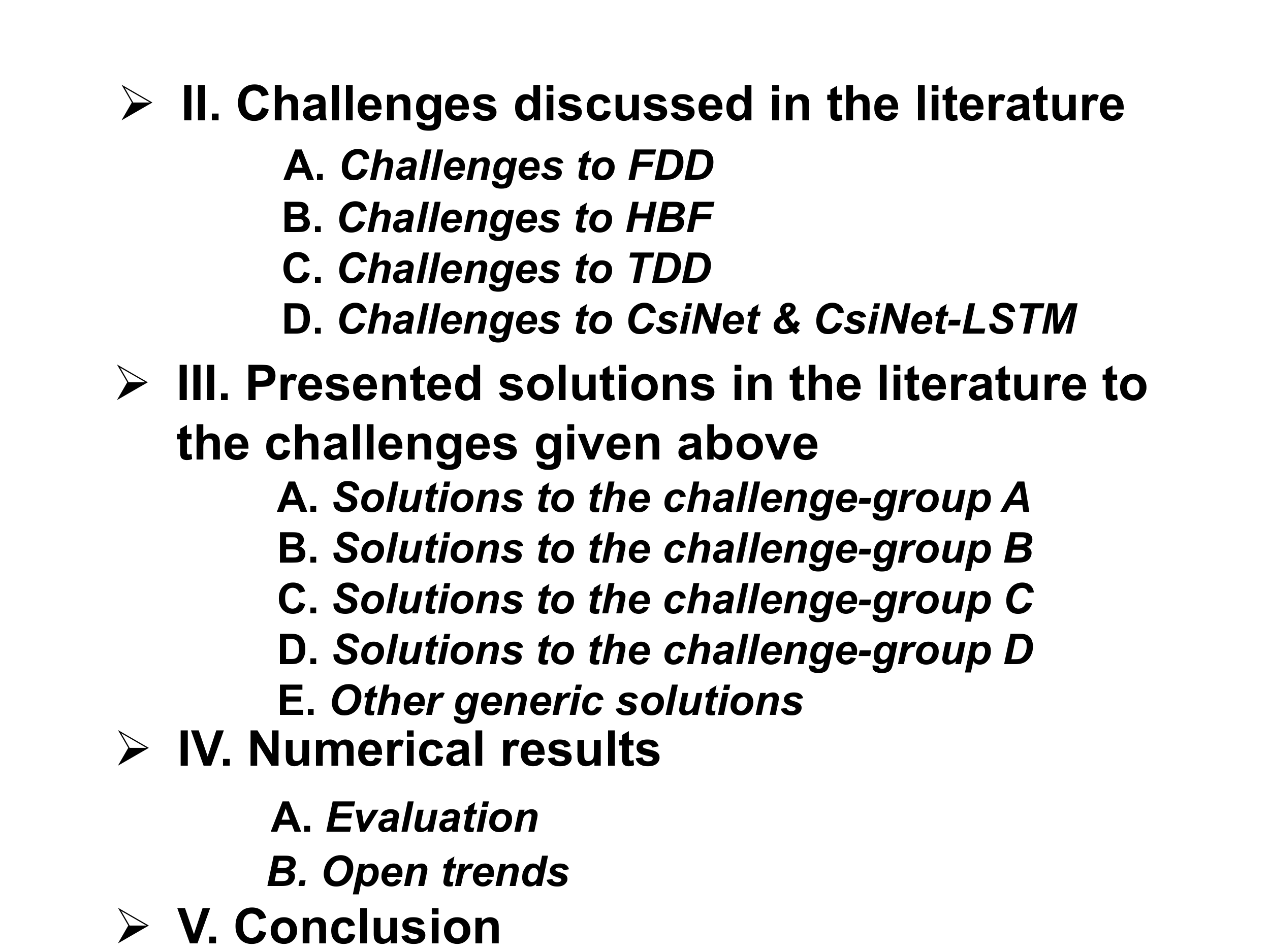}} 
\caption{Structure of this paper.}  \label{F0}
\label{fig:EcUND} 
\end{figure} 

\subsection{Motivations and contributions}
More acceptable versions of pilot-data trade-off, feedback overhead, novelty, optimality, algorithms and architectures are still required. These open trends motivate us to present this survey in relation to DLg based channel modeling/prediction in MaMIMO. In this state-of-the-art survey, we review the work performed in the literature. All the challenges that have been or not been considered are fully discussed. More specifically: 
\begin{itemize}
\item We actualise a snapshot of the major methods and solutions included so far. This survey is the sole one which is presented in the context of DLg based channel modeling/prediction in MaMIMO - something that has not been performed until now.
\item We provide technical details of the relative techniques covered along with the topic, categorising them into 4 totally different groups. 
\item We also consequently propose 7 open trends in the context of the following phrases: (\textit{i}) lack of Q-learning in MaMIMO detection - for which we consider a possible solution to the saddle-point in relation to the pilot-data trade-off for a \textit{Stackelberg game} based scenario; (\textit{ii}) frequency synchronisation and DLg-based MaMIMO detection; (\textit{iii}) pilot contamination and imperfect CSI in DLg-based MaMIMO detection; (\textit{iv}) more acceptable algorithms and architectures w.r.t. the principle of "complexity vs. accuracy" in DLg-based MaMIMO detection; (\textit{v}) lack of information theoretic work in DLg-based MaMIMO detection; (\textit{vi}) ADC impairments in DLg-based detection; and (\textit{vii}) challenges in relation to the physical reciprocity between UL and DL in practice in DLg-based transceivers.
\item Sufficient discussions are provided with the goal of useful comparisons. 
\end{itemize}

\subsection{Organisation \& preliminaries }
\subsubsection{Organisation}
The rest of the paper is organised as follows. Challenges discussed in the literature in relation to MaMIMO and in the context of DLg based channel modeling/estimation are evaluated in Section II in-depth. Subsequently, the recently proposed solutions are discussed in Section III. Evaluations and more discussions are given in Section IV, after which, conclusions are listed in the last section. The trend of this paper is given in Fig. \ref{F0}.

\subsubsection{Preliminaries}
Acronyms used throughout the paper are listed in Table \ref{table1}

\captionof{table}{{Acronyms.}}
 \label{table1} 
\begin{tabular}{p{6.71cm}p{0.1cm} }
\textsc{\textbf{Term}}        &  \textsc{\textbf{Acronym}}  \\
Base-station & BS    \\    
Millimeter-wave & mmWave  \\   
Downlink & DL     \\    
Uplink & UL \\ 
 Deep-learning & DLg     \\    
Bandwidth & BW \\  
Deep-neural-network & DNN     \\   
 Fifth-generation & 5G \\ 
 Time-division-duplexing & TDD     \\
    Frequency-division-duplexing & FDD \\  
Orthogonal-frequency-division-multiplexing & OFDM     \\    
Massive multi-input multi-output  & MaMIMO \\ 
 Convolutional-neural-network & CNN     \\    
 Convolutional & Conv     \\    
Network & Net \\  
CSI-Network & CsiNet \\
Long-short-term-memory & LSTM \\
Channel-state-information & CSI    \\
Hybrid beam-forming & HBF    \\
Compressed-sensing & CS    \\
User equipment & UE   \\
Minimum-mean-square-error  & MMSE    \\
Mean-square-error  & MSE    \\
Radio frequency  & RF \\
Direction-of-arrival & DoA \\
Dimension & D\\
In-phase-and-quadrature & IQ \\
Ultra-dense-network & UDN\\
Maximum-likelihood & ML \\
Non-line-of-sight & NLoS\\
Line-of-sight & LoS\\
Signal-to-interference-noise-ratio & SINR \\
Signal-to-noise-ratio  & SNR \\
Analog-to-digital-convertor & ADC\\
Probability-mass-function  & PMF \\
Supervised-learning  & SLg \\
Unsupervised-learning  & ULg \\
Reinforcement-learning & RLg \\
Evolved-node-B & eNB\\
\end{tabular}

\section{Challenges discussed in the literature}

In this section, we overview the challenges discussed in the literature in terms of DLg in MaMIMO systems, prior of which we briefly go over DLg.

A simple deep-neural-network (DNN) for MaMIMO transceivers is depicted in Fig. \ref{F1}. The information-theoretic philosophy behind of DLg also physically originates from the following rule \cite{R3R3}: \textit{The least-upper-bound for the compression rate must be found according to which the reconstructed data ($\hat{\mathcal{X}}$) must be most informative w.r.t. the original one ($\mathcal{X}$, and the real-output $\mathcal{Y}$).} This is in correspondence with the information-rate-distortion problem of 
\begin{equation*}
\mathop{{\rm min}}\limits_{PMFs} {\rm \; }\left \{ \mathcal{I}(\mathcal{X};\hat{\mathcal{X}}) \right \},
\end{equation*}
subjected to
\begin{equation*}
dist(\mathcal{X},\hat{\mathcal{X}})=\mathcal{I}(\mathcal{X};\mathcal{Y})-\mathcal{I}(\mathcal{Y};\hat{\mathcal{X}}) \le \gamma_1,
\end{equation*}
should hold w.r.t. the threshold $\gamma_1$, where $\mathcal{I}(\cdot;\cdot)$ and $dist(\cdot,\cdot)$ are respectively the mutual-information and distance. This is finally equivalent to 
\begin{equation*}
\mathop{{\rm max}}\limits_{PMFs} {\rm \; } \mathcal{I}(\mathcal{Y};\hat{\mathcal{X}})-\gamma_2\mathcal{I}(\mathcal{X};\hat{\mathcal{X}}),
\end{equation*}
w.r.t. the Lagrange multiplier $\gamma_2$, and the required optimum-values for the probability-mass-functions (PMFs). In other words, the principle of \textit{information-Bottleneck} \cite{R3R3} principally interprets: \textit{how to find the least number of the hidden layers according to which our reconstructed data is most informative w.r.t. the original one}. 

DLg is mainly classified to two essential categories:
\begin{itemize}
\item Supervised-learning (SLg): where the features are labeled, with a high accuracy and a high complexity.
\item Unsupervised-learning (ULg): in which the features are clustered, with a less accuracy and a less complexity compared with SLg.
\end{itemize}
In addition to the two categories expressed above, there is also semi-supervised-learning, which is widely called hybrid-learning. In this approach, some features are labeled, while clustering the remaining features. Finally, there is another learning approach called reinforcement-learning (RLg) in which \textit{policy-learning} mostly plays the vital role.

\subsection{Challenges to frequency division duplexing (FDD)} In FDD, the channel reciprocity is not satisfied due to the fact that the forward and reverse links generally have highly uncorrelated channels \cite{R4R4}. In FDD, however, it is illustrated that the correlation of shadowing also exists between UL and DL \cite{R5R5}. There also seems to be a small multipath correlation between UL and DL channels in MaMIMO systems. However, the weak situation for the phase correlation degrades the phase recovery. In polar coordinate, additionally, uniform phase quantization results in quantisation error.

There seems to be one totally important criterion which helps in the overhead-reduction interpretation as follows \cite{R6R6}. There is a small angular spread in relation to the channels between the base-station (BS) and User-equipments (UEs). This proves a sparsity in the angular domain. In FDD MaMIMO transceivers, compressed-sensing (CS)-based CSI feedback mechanisms\footnote{To sparsify CSI.} are emerging in order to relax the channel dimension \cite{R7R7}. This is fulfilled with the aid of employing the sparse structures of CSI e.g., CSI's temporal correlation, CSI's spatial correlation and the sparsity-enhancing basis for CSI etc. The sparsity of CSI only holds for a few special models; something that cannot be generalised to a widely-accepted method for the case of model mismatch. As discussed in \cite{R1R1}, the majority of the models in relation to time correlation is the block fading one. By this, the CSI dimensionality is relaxed by a factor of the blocklength. Also, the time correlation in slow-fading channels plays a key role \cite{R8R8}. W.r.t. a certain threshold in relation to the reconstruction error\footnote{Relating to the distortion/loss function.} the relative methods reuse the previously retained CSI for the subsequent CSI reconstruction. However, the reused information is hard to be updated in a real-time manner. For the fast-fading channels the problem is going to be worse since the resolution is degraded resulting in the failure of the feedback-overhead reduction.

It is required to leverage the CSI linear correlations in the spatial and time domains as well as the frequency one. The literature mainly concentrates on the spatial correlation by transforming the CSI to the angular domain. Moreover, the time correlation is physically modeled by a Gaussian-Markov process. Some recent work mentioned in \cite{R9R9} in which an efficient orthogonal-space-time-frequency coding scheme was technically proposed for OFDM based systems. The aforementioned scheme talked about how to employ the CSI correlation in the frequency domain which lies in the sparse multipath component. Additionally, the linear correlation among collocated antenna sets reduces the CSI overhead in the spatial domain. However, the distance between two remote BSs degrades this assumption.

Additionally, five fundamental bottlenecks inherently \cite{R4R4, R6R6, R7R7, R10R10, R11R11, R12R12} exist in the CS-based techniques which are mentioned in the following. 

\begin{figure}[t]
\centering
{\includegraphics[trim={{1mm} {0 mm} {1 mm} {0mm}},clip,scale=0.540462]{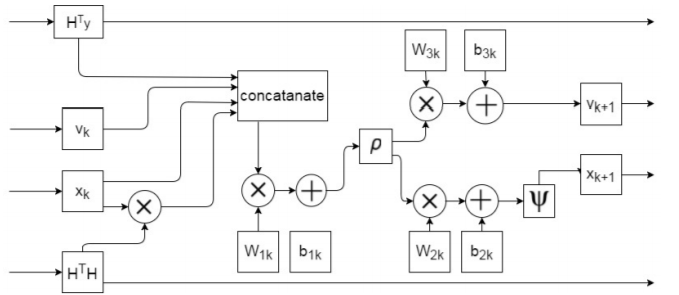}} 
\caption{DetNet proposed in \cite{R15R15}.}  \label{F2}
\label{fig:EcUND} 
\end{figure} 

\begin{itemize}
\item Firstly, the CSI matrix is not precisely sparse in any specific domain. In other words, channels are considered sparse, however, this assumption is in some conditions somehow illogical. 
\item Secondly, the computational overhead will any way be high since CS-based algorithms perform time-consuming in an iterative manner (slow recovery). 
\item Thirdly, the variations in the sparse mmWave channel matrix between the neighbors are totally tiny and highly-complicated to be modeled \cite{R6R6}. 
\item Fourthly, random projection is used in CS instead of fully employed channel structures \cite{R11R11}. Indeed, random matrix theory applicable in CS is also off-topic in reality specifically when the dimension of compressed measurements is not sufficient. 
\item Fifthly, either Gaussian or Bernoulli distributions are often considered which are not optimal for all channel models in practice \cite{R12R12}.
\end{itemize}

In FDD, vector quantization or/and codebook-based approaches may be useful for some conditions \cite{R11R11}. However, the highly complicated structure of the code-book w.r.t. a high size of antenna arrays, and the quantization errors result a vast range of bottlenecks to CSI \cite{R4R4, R8R8}.

Two-stage precoding scheme is widely used for FDD MaMIMO systems aimed at relaxing CSI feedback-overhead. The inner precoder employs a zero-forcing equalisation w.r.t. local CSI. The UEs are also clustered by the outer precoder w.r.t. the conformity of the eigen-space of the auto-correlation matrix related to the UEs' DL channel \cite{R4R4}.

In MaMIMO systems, w.r.t. the pilot-length $L_s$, barely does the condition of $L_s \ge N_t$ hold \cite{R13R13}. This is due to three main reasons which are expressed as follows. First, w.r.t. the aforementioned condition, a huge amount of time is necessary for the pilot transmission. This consequently leads to a less spectral efficiency w.r.t. the data-pilot trade-off. Secondly, the complexity rises as $L_s$ rises. Finally, the condition cited above may be impossible in some circumstances, since $L_s$ cannot be larger than the uncontrollable criterion of the channel-coherence-time. On the other hand, the condition $L_s < N_t$ for the MaMIMO channel estimation is somehow impossible in practical situation due the following reasons. It is, in fact, impossible to design the pilot matrix in the sense that its row vectors would be able to be orthogonal. However, this orthogonality is required for interference mitigation among the pilot sequences transmitted from various antennas. The aforementioned impossibility indicates $L_s \nless  N_t$. Secondly, there is no guarantee for the optimality of the MMSE channel estimator if $L_s < N_t$ holds. This comes from a non-convex nature of the original optimisation problem.

\subsubsection{Challenges to the DLg-based CSI feedback}
DLg-based CSI feedback also still shows inefficiency to decrease the occupancy of UP-BW resources, even though it tackles a lot of obstacles \cite{R7R7}. An optimisation over the usage of the UL-BW is still highly regarded since the chief focus of the literature for DL-based CSI feedback in FDD MaMIMO systems is on feedback reduction. In the training phase, pre-marked data are fed into the Net in order to find the most optimised version of the Net-connection weights. Towards this end, the gradient descent method is widely used. However, the gradient will vanish when the Net-layers are high \cite{R14R14}; something that results in increasing the training time and even relevant-information drop-out. Additionally, the performance of the detection-Net called DetNet \cite{R15R15} (shown in Fig. \ref{F2}\footnote{in which $W$ and $b$ are respectively weights and biases.}) is, with the decrease of the number of antenna sets, worse\footnote{Although DetNet has an acceptable robustness and flexibility, since it can adapt to an extended SNR regime and for various channels.}. Furthermore, DetNet fails if the number of transmitter's antennas is close to or larger than the number of receiving antennas.

\subsection{Challenges to hybrid beam-forming (HBF)} For realistic MaMIMO transceivers, exploiting fully digital baseband counterpart, that is, the use of the same number of radio frequency (RF) chains with the size of the antenna sets is unpractical. In other words, in mmWave MaMIMO transceivers, it is somehow impossible to assign each antenna to the RF chain. The main problem is indeed the energy consumption as well as the hardware cost. In order to guarantee a widely acceptable trade-off between system performance and hardware cost, the principle of HBF is considered. This type of novel beamforming divides the beamforming operations between the analog and digital domains. HBF exploits a low-dimensional digital precoder in the subsequence with a high-dimensional analog-precoder. Therefore, a sufficient-and-necessary number of antennas with much fewer RF chains are assigned. 
The signals are phase shifted in the RF\footnote{Since the relistic RF power harvesting counterpart includes only rectifiers, no ability of the RF-to-baseband conversion (baseband processing generally speaking) exists. This causes an impossible actualisation of a realistic energy
receiver.} using analog phase shifters prior of which they are digitally precoded in the baseband\footnote{These HBFs are twofold namely fully-connected architectures and sub-array-connected ones.}. This issue enforces the 5G Net operators to find a significantly more suitable solution instead of exploiting simple traditional time-division-duplexing (TDD)/FDD techniques (prevalently used in the previous generations). This is because when the receiver is equipped with a limited number of RF chains, channel/direction-of-arrival (DOA) estimation is totally difficult \cite{R16R16}, \cite{R17R17}, \cite{R18R18}, \cite{R19R19}. In other words, in a MaMIMO transceiver equipped with a large number of unknown channel parameters but a small number of RF chains, a huge amount of time is required for training phase, resulting a long delay. In this delay time, channel may be varied. Furthermore, mmWave channels is of a purely limited scattering-based nature, resulting in a specific angular sparsity for which CS may be required. In vehicular communications \cite{R20R20} in which the mobility of devices creates time-varying channels.

\begin{figure}[t]
\centering
{\includegraphics[trim={{0mm} {0 mm} {0 mm} {0mm}},clip,scale=0.400462]{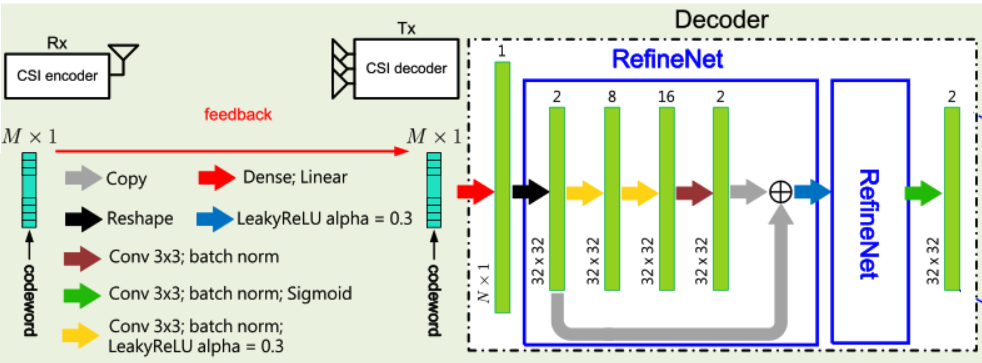}} 
\caption{CsiNet proposed in \cite{R11R11}.}  \label{F3}
\label{fig:EcUND} 
\end{figure}

\subsection{Challenges to TDD} The same orthogonal pilot sequence among cells can be highly probably reused in cellular systems which results in "pilot contamination" \cite{R21R21, R22R22, R23R23}. Existing work lack in relaxing this issue which is owing to the fact that the pilot assignment problem has a large search space over cells which makes a non-convex optimisation problem. One of the best solutions, as discussed in \cite{R22R22}, to alleviate pilot contamination is determined as follows, while being knowledgeable of the pilots in adjacent cells. One may assign random pilots in every cell and estimate both the in-and-out-of-cell user channels at each BS.

Nonlinear hardware counterpart such as signal amplifiers and analog filters highly complicates the calibration \cite{R24R24}. In practice, the analog front-end circuitry for transmission and reception at the BS and the UEs, generally speaking, does not have any reciprocity in TDD. The process of channel calibration contains two steps: (\textit{i}) how to predict the calibration coefficients between the UL and DL channels; and (\textit{ii}) a compensation over calibration.

\subsection{Challenges to CSI-Network (CsiNet) \cite{R11R11} and CsiNet-long-short-term-memory (CsiNet-LSTM) \cite{R8R8}} 

In CsiNet, the values $S1$, $S2$ and $S3$ stand respectively for the length, width, and number of feature maps. The first layer of the encoder is a convolutional (Conv) one for which the real and imaginary parts of $\mathbf{H}$ are the input. In order to produce two feature maps, this layer exploits kernels with dimensions of $3  \times  3$. In the subsequence with the Conv-ayer, the feature-maps are reshaped into a vector. This utilise a fully connected layer to create the code-word vector $s$ of size $M$. The first two layers play as encoders. CsiNet translates the extracted feature maps into a code-word in contrast to CS. The first layer of the decoder is a fully connected layer. The input is $s$ according to which the outputs are two matrices of size $N_r  \times    N_t$, which stand for the reconstructed version of the real and imaginary parts of the channel matrix $\mathbf{H}$.

\begin{figure}[t]
\centering
{\includegraphics[trim={{1mm} {0 mm} {1 mm} {4mm}},clip,scale=0.462]{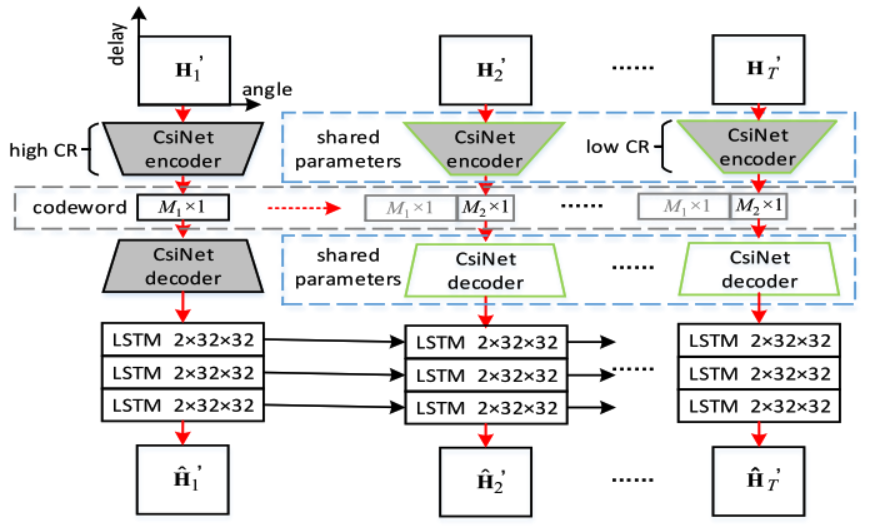}} 
\caption{CsiNet-LSTM proposed in \cite{R8R8}.}  \label{F4}
\label{fig:EcUND} 
\end{figure}

As discussed in \cite{R7R7}, \cite{R25R25}, the CsiNet \cite{R11R11} (shown in Fig. \ref{F3}) and CsiNet-LSTM (shown in Fig. \ref{F4}) are not applicable in reality in time-varying channels. This is owing to the negligence of time correlation since the CSI is independently reconstructed in the CsiNet module. This is indeed because the employment of linear fully-connected Nets are unsuitable for applying the temporal correlations. Finally, spatial correlation between antenna sets was not considered in CsiNet. 

\textsc{\textbf{Remark 1:}} \textit{There seems to be two classes in the existing work related to the wideband
mmWave channel prediction: the time-domain prediction and frequency-domain
one. The first class concurrently pridictes all channel taps,
inversely, the second one independently predicts individual subcarriers. It is also interseting to note that both classes result
similar performances if the principle of angular sparsity is applied. However, 
heavy computational complexity occoures for the both classes if CS techniques are employed.}

\section{ Presented solutions in the literature to the challenges given above}

\subsection{Solutions to the challenge-group A}

In \cite{R1R1}, a remote CSI-feature-extraction scheme was proposed applying DLg for FDD MaMIMO systems.

In \cite{R2R2}, a 3-level DLg-based scheme (called OCEAN) was technically proposed. This aimed at analysing the archived data for online CSI estimation containing a 2D CNN Net, a 1D CNN Net, and an LSTM-Net. Specifically, in the first stage, the 2D CNN Net is used to construct the frequency-feature vector from CSI images. In this level, since the CSI images are high-dimensional tensors, a flatten-layer applies a concatenation procedure by which these tensors are alleviated into 1-D vectors.  In particular, CSI is actualised with an integration of the influences stem from path loss, diffraction, scattering, fading, shadowing, frequency band, location, time, temperature, humidity, and weather.  Indeed, the spatial-temporal relevance (correlation) of the CSI resulted in proposing a DL-based scheme which is an integration of a CNN and an LSTM-Net. To validate OCEAN's efficacy, the authors tested OCEAN's performance by considering four typical case studies. The authors examined: two outdoor environments, viz., a parking environment and outside a building; and two indoor environments, viz., a workroom and a building. The stability of the framework was also enhanced via a bi-level offline\footnote{Training phase with a low underfitting error.}-online\footnote{Testing phase with a low overfitting error.} learning procedure. The offline training step for analysing of the historical data was theoretically characterised in the offline step, subsequently, in the online training phase, the estimated and recorded
CSI were added to each other. When OCEAN receives a CSI-estimation request, it initially collects real-time data of all criteria\footnote{Side-information in in information-theory, as the relevant one but not the perfect one.}. OCEAN subsequently according to the side information conducts the
CSI estimation. The side-information is three-fold consisting 
Frequency band, Location and Time. Firstly, the atmosphere density is totally various in different seasons and even changing day-by-day. Moreover, Temperature can affect on the atmosphere, which has a knock-on impact on the scattering and fading of radio-signal propagation. Humidity in the context of the weather, specifically, rain degrades the power of the transmitted radio signals through frequency bands.

A new proposition was theoretically characterised in \cite{R4R4} in which the number of feedback
sets can even be smaller than the size of the channel
matrix (without loss of optimality and flexibility). 
The authors' DLg work
efficiently guarantees up to 1000x reinforcement in computational complexity,
efficiently ensuring high estimation quality even for a finite data sets.
A new-and-brilliant low-rank model of the double-directional
MaMIMO channel was perfectly propose to efficiently conduct a DLg-based estimation framework. The main idea fundamentally is: (\textit{i}) how to efficiently model the
DL channel with a few number of paths in the context of a low-rank channel;
and (\textit{ii}) how to efficiently reconstruct such channels at the BS with the help of a small
range of scalar measurements sent from the UEs.
In this scheme, UEs only
principally apply random compression over the received pilot signals subsequently
sending the compressed image back to the BS. 
In order to totally solve the low-rank model at the BS, 
two DLg based approaches were theoretically proposed. Aside from many existing DLg-based
techniques which train algorithms/classifiers,
a complex-valued low-rank model is efficiently learned/trained in their scheme. The logic behind of this scheme hence is how to basicaaly learn/train
the nonlinear inverse relevance/proportion of the feedback samples to the true (complex)
low-rank channel.

In \cite{R5R5}, a DLg-based CSI feedback mechanism was theoretically proposed for limited
feedback and bi-directional reciprocal channel characteristics.
The BS efficiently applies the available uplink (UL)
CSI to totally reconstruct the unknown downlink (DL) CSI from low
rate user feedback. 
This was theoretically characterised to relax the CSI
feedback payload basedrelying upon the multipath reciprocity. 
The absolute value of real/imaginary parts of
the CSI data sets as well as the bi-directional correlation
of the relative magnitudes are the two totally central keys.
Both UL
CSI and DL feedback are the input of the DLg Nets for
decoding the most informative and efficient feedback code-word. The novelty of this
work was an efficient use of the fact that the bi-directional channel correlation
ensures a fundamentally performance reinforcement for DL
CSI prediction in FDD systems with limited UL feedback.
In order to perfectly relax the feedback BW, the CSI
quantisation error is efficiently restricted in the authors' work. This was perfectly actualised via magnitude dependent phase
quantisation in which CSI coefficients with smaller
magnitude are in accordance with the less phase quantisation. The
Evolved-node-B (eNB) subsequently restores the quantified phase efficiently using the
reconstruct magnitude. In this scheme, the
code-word lengths may basically vary w.r.t. the magnitudes, hence, the mean
code-word length depends mainly upon the distribution of CSI magnitude.

In \cite{R6R6}, an efficient and deterministic UL-to-DL mapping
function is considered while the position-to-channel mapping is theoretically bi-jective.
According to the universal approximation theorem, this function is estimated. 
After offline-training, the SCNet shall efficiently estimate the DL
CSI w.r.t. the UL CSI with neither
the DL training nor the UL feedback. The authors' model essentially divides an image into several sub-images technically aimed at relaxing
the offline-training and online-testing latency.

In order to further reduce the UL-BW-occupation, DLg and superimposed coding for CSI feedback were efficiently integrated in \cite{R7R7}. The DL CSI was initially spanned and subsequently superimposed over/on UL user data points towards the BS. Towards this end, a multi-task DLg architecture was technically proposed at BS, with MMSE-based interference mitigation, basically aimed at efficiently recovering the DL-CSI and UL-US.

The authors in \cite{R8R8} found an efficient extension for CsiNet with LSTM to find an enhancement in compression-ratio and recovery quality trade-off. This framework strongly provides an efficient alleviation to the parameter overhead in FDD schemes. 

The Cramer-Rao lower bound of remote CSI inference was totally explored given the local CSI in \cite{R9R9}. 
The achievable CRLB proves the relevance between the CSI-inference and some totally important system-parameters such as antenna array size. W.r.t. a low inference error the efficiency of the authors' DNN-based CSI inference is validated.

A Conv-LSTM-Net driven DLg approach in \cite{R10R10} was technically proposed for the direct prediction of the DL-CSI from the UL-CSI. This was completely performed with the goal of efficiently tackling the complexity and feedback-overhead of FDD-MaMIMO. In the proposed bi-level framework, the feature-extractor module in the first stage learns the spatial correlation as well as the temporal one between the DL-CSI and the UL-CSI. Subsequently, in the second phase the prediction module maps the extracted features to the reconstructions of the DL-CSI. The authors' simulations showed an acceptable efficiency, specifically in the time domain.

A developed CSI sensing (or encoder) and recovery (or decoder) Net was technically proposed called CsiNet in \cite{R11R11}. At the encoder, instead of applying a random-projection mechanism, CsiNet is efficiently trained, in the offline-training phase, a transformation from original channel coefficients
to compress code-words at UE. CsiNet is also non-iteratively trained an inverse transformation from code-words to original channels at the BS.

In \cite{R12R12}, a novel Dlg based method to efficiently estimate compressed DL CSI feedback for FDD oriented MaMIMO systems was proposed as well. Indeed, DLg was fundamentally used to further improve the CS method.

\cite{R13R13} considered a case in which the
pilot length is basically smaller than the number of transmit antenna sets.
A two-stage prediction mechanism was technically proposed. In the first step, the pilot in addition to
the channel estimator are concurrently DLg-based designed
in the second step, the accuracy of channel estimation
is also iteratively improved.

By a half amount of connectivity, being guaranteed by sparsity, a low-complex DLg Net was technically proposed in \cite{R14R14}. For this Net, a less amount of input-data is basically required with an acceptable overall-performance.

\begin{figure}[t]
\centering
{\includegraphics[trim={{42mm} {99 mm} {56 mm} {35mm}},clip,scale=0.4]{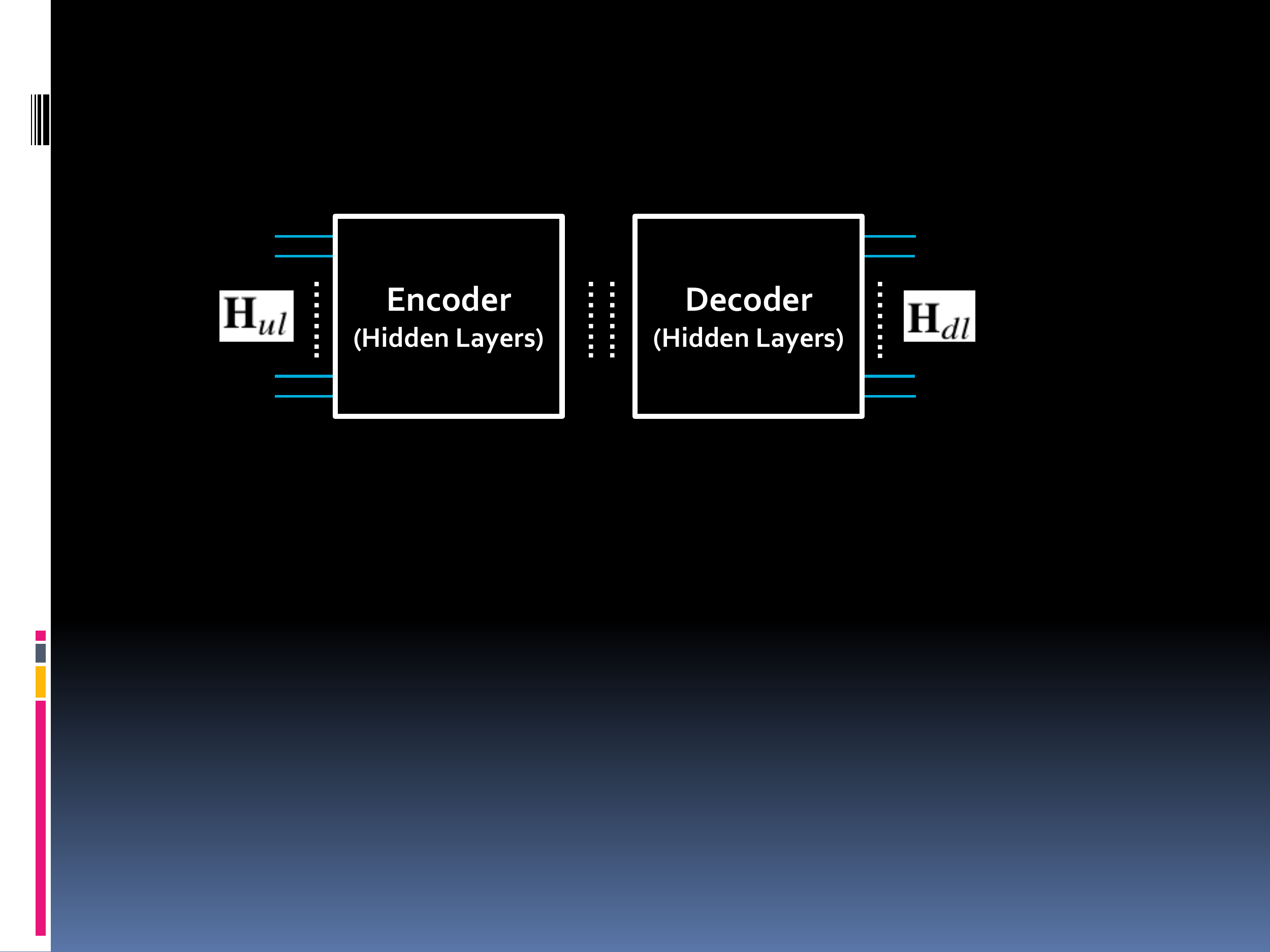}} 
\caption{A DNN for MaMIMO proposed in \cite{R24R24}: How to estimate the DL channel matrix from the UL one.}  \label{F5}
\label{fig:EcUND} 
\end{figure} 

\subsection{Solutions to the challenge-group B}
In \cite{R16R16}, a learned de-noising based approximate message passing mechanism was theoretically proposed. This scheme efficiently learns the overall structure of the channel and efficiently estimates the channel. In the simulation, the scheme is implemented with MatCovNet as a toolbox in MATLAB.

A spatial-frequency CNN driven channel prediction was essentially realised w.r.t. the spatial correlation was well as the frequency one in \cite{R17R17}. This was efficiently realised where the corrupted channel coefficients at neigbour subcarriers are the input of the CNN. Subsequently, a case of the temporal correlation in time-varying channels is efficiently deployed.
The authors' proposed solution can perfectly guarantee a sub-optimal throughput near-optimally close to the ideal MMSE estimator. The design of MMSE hardly is it possible to be efficiently performed in practice. The authors' work is additionally robust to various propagation aspects.

A low-complexity DLg oriented DOA estimation procedure was technically proposed in \cite{R18R18} for HBF in MaMIMO systems with a uniform circular array at the BS. The proposed method is bi-level. In the first stage, the received signal vector is initially input into some small deep feed-forward Nets. These inputs are supposed to be efficiently trained in an offline manner. Consequently, in the second level, a set of candidate angles are produced among which the optimal one is selected. Compared to the conventional maximum likelihood (ML) method, the proposed DOA estimation method totally outweighs, from a complexity point-of-view. 

In \cite{R19R19}, a new time-domain channel
estimation mechanism for hybrid mmWave was theoretically characterised. In this scheme, the
well-known angular sparsity is efficiently used as well as the delay-domain sparsity. Both channel sparsity
(in the delay domain) and the angular sparsity are used, to efficiently relax
the training-overhead as well as computational complexity. 
A novel time-domain channel estimator was indeed perfectly proposed by innovatively considering such double sparsity. 

\begin{figure*}[t]
\centering
\subfloat[Cell-spectral-efficiency (FDD) ]{\includegraphics[trim={{7 mm} {9 mm} {6mm} {1mm}},clip,scale=0.28]{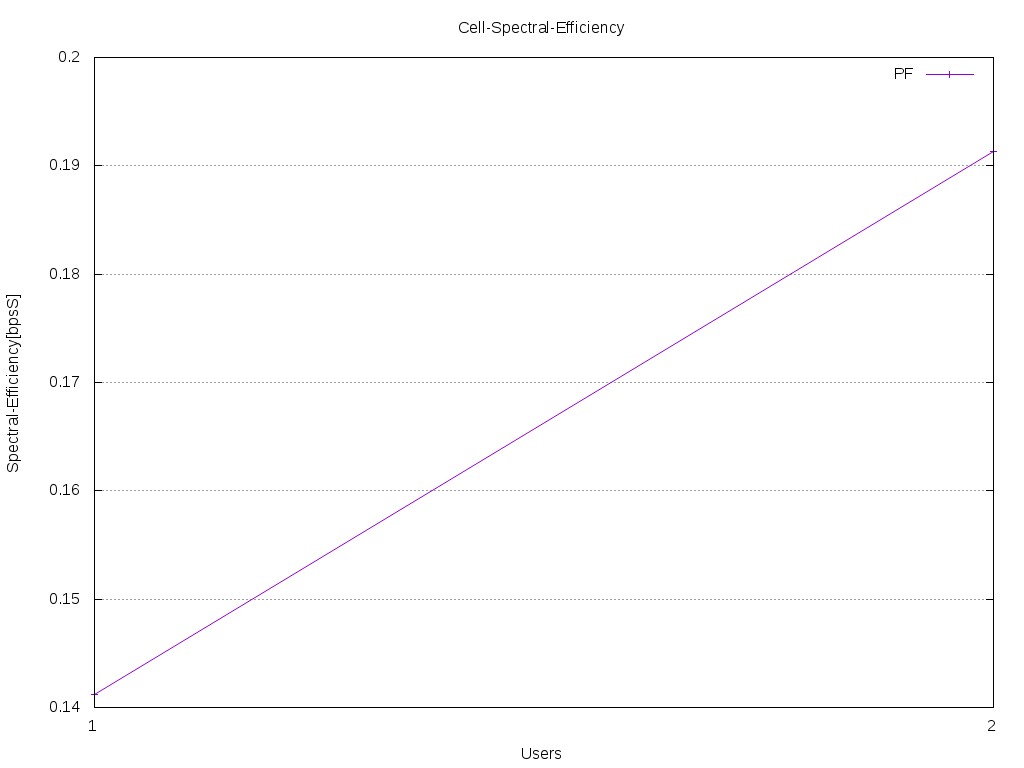}} 
\subfloat[Cell-spectral-efficiency (TDD) ]{\includegraphics[trim={{7 mm} {9 mm} {6mm} {1mm}},clip,scale=0.28]{Spectral-Efficiency_Spec-Eff.jpg}} 
\caption{FDD vs. TDD.} \label{F6}
\label{fig:EcUND} 
\end{figure*} 

A novel DLg-based framework was technically proposed which collects in-phase-and-quadrature (IQ) samples of the IEEE 802.11p transmission applying CSI-features extraction algorithms in \cite{R20R20}.
This can also efficiently estimate the CSI and received signal-levels.
\subsection{Solutions to the challenge-group C}
In \cite{R21R21}, a super-resolution DOA estimation framework was essentially proposed exploiting DLg for TDD based MaMIMO systems.

How to efficiently re-construct a sparse signal from a few noisy
linear constructions was basically explored in \cite{R22R22}. 
Two new DNN
architectures were technically deployed which can efficiently cope with estimation errors over
layers. A joint DLg mechanism was technically proposed which can efficiently learn
the linear transforms as well as the nonlinearities.

In \cite {R23R23}, the authors efficiently proposed a novel performance enhancement in cellular Nets by relaxing pilot contamination through a novel SLg of the relevance of pilot assignment in addition to the users' location patterns. The input features and output labels are respectively users' locations in all cells and pilot assignments. Pre-trained optimal pilot assignments with given users' locations, in particular, were efficiently guaranteed via an exhaustive search procedure.

A novel design (shown in Fig. \ref{F5}) for relaxation over the effect of the spatially correlated channel 
was theoretically characterised in \cite{R24R24} by efficiently creating 
the 3-D correlated channel models. 
The main goal was to do channel calibration between the UL and DL channel coefficients.
The haphazardly
predicted UL channel was efficiently used to calibrate the DL one, which is
totally unobservable within the UL transmission phase. 
The authors' designed Dlg-based work is efficiently applicable even for 
FDD MaMIMO
systems with nonlinear hardware transceivers.
\subsection{Solutions to the challenge-group D}
In \cite{R25R25},
a DLg-based CSI compression feedback mechanism for multi-user MaMIMOs for FDD maMIMO systems taking into account the spatial correlation of MaMIMO channel was proposed. This respectively uses Bi-LSTM and Bi-Conv-LSTM Net
in order to efficiently reconstruct the CSI for respectively single-user and multi-user samples. 
In decompression process, initially, a fully connected
layer is basically used to increase the compressed CSI data
dimension to the dimension before compression.
\subsection{Other generic solutions}
In \cite{R26R26}, the problem of a localised prediction of the traffic-load at the ultra-dense-networks (UDNs) BS, that is, the eNB was essentially explored. This is due to the challengeable complexity and dynamicity of traffic-flow (e.g. the non-casual manner of buffer-occupancy status) in UDNs in MaMIMO. The localised prediction was efficiently actualised w.r.t. the output of the LSTM which perfectly indicates the possible occurrence of the congestion. The authors' framework has a better packet-loss-ratio and overall-throughput. 

\cite{R27R27} technically proposed\footnote{Since system-capacity improvement and service coverage degradations are fundamentally in the expense of each other (called capacity-coverage tradeoff).} a new solution to the aforementioned issue in MaMIMO, considering inter-cell interference conditions. The proposed scheme was a group alignment of user signal strength which perfectly guaranteed the user scheduling in MaMIMO. The Quality of Service criterion was also the minimum signal-to-interference-plus-noise ratio (SINR). The variance of signal strengths of multiple adjacent users was supposed to be efficiently supported. In order to dynamically guaranteeing this, the authors' method was DLg driven.

DLg technique was basically utilised to efficiently train the user-mobility (channel dispersion from a frequency-domain point-of-view) in \cite{R28R28}. This work totally proposed a 3-level DLg approach in the context of the adversarial RLg workflow. In the first step, the DNN is basically trained to efficiently produce realistic user mobility patterns. Accordingly, the second DNN in the second phase is fundamentally trained to generate relevant antenna diagram. Finally, the third DNN in the last stage estimates the efficiency of the generated antenna patterns. The first and foremost beneficial feature of the proposed approach was the fact that this is self-trained while it is not necessary to have a large training data sets.

A DLg technique was efficiently applied to reduce the time of beamforming in \cite{R29R29}.
The DL transmission for full-dimension MaMIMO transceivers was basically discussed over correlated Rician fading channels. 

A new-and-efficient channel-sounder-architecture was technically proposed in \cite{R30R30}, in order to essentially estimate the CSI at different frequency bands, antenna geometries and propagation environments. This was totally fulfilled for a multi-carrier MaMIMO system. The authors achieved an accuracy better than $75$ cm for line-of-sight (LoS) compared to the previous literature, as well as the case of non-line-of-sight (NLoS) for FDDs.

The authors in \cite{R31R31} efficiently examined the usability of DNNs for MIMO user positioning relying upon OFDM complex channel samples. The authors efficiently proposed a DLg-based framework for MaMIMO-OFDM transceivers for which, in contrast to other indoor positioning systems, it was required no piloting overhead in NLoS scenario. Since gradient descent method requires an enormous range of data-sets in the training phase, a two-step training procedure was consequently proposed. In the first level, simulated LoS data-sets were efficiently trained, subsequently, the measured NLoS positions in the second phase was basically tuned. This procedure efficiently reduces the required recorded training-positions as well as the complexity of data reconstruction.

In order to deal with the feedback overhead, a remote CSI inference method was technically proposed in \cite{R32R32}. This was essentially realised with the aid of probing the channels occupied by a source BS and inferring the CSI of target BSs at totally different sites. This work essentially was a generalisation of the previous research which chiefly concentrates on the usage of the CSI-linear-correlations of neighbor antenna arrays. The scheme was a DLg one to efficiently explore non-linear dependence among remote CSI. The existence of such cross-BS CSI dependence was perfectly proven with the aid of the calculation of the mutual information between remote CSI, and the Cramer-Rao lower-bound of remote CSI inference.

In \cite{R33R33}, Keras is efficiently utilised to construct and process the DNN part. A DLg based HBF was theoretically proposed with a well-accepted complexity which basically receives
structural information 
within the training.

Trainable Projected Gradient-Detector as a DLg-
aided iterative decoder was technically proposed in \cite{R34R34}.
A new application of data-driven tuning to MaMIMO
detectors was indeed efficiently conducted.

An ULg based pilot power allocation was essentially realised in \cite{R35R35}. A
DLg is indeed efficiently designed to map the input as large-scale fading channel data sets to the output as the pilot power allocation vector where the sum MSE is the loss function. The proposed pilot power allocation scheme is totally implementable to the case of large number of users. This is essentially because the ground truth is not mandatory
in ULg in comparison with the SLg.

A joint data-pilot power-control problem in the context of a SLg one (aimed at efficiently dealing with the non-convexity) was basically explored in \cite{R36R36}. This was essentially realised where the transmit-power elements is efficiently estimated through an input of large-scale fading channel data sets. The loss function was also the weighted MSE.

A brilliant proposition of mapping the channel coefficients at a given set of antenna arrays and a given frequency band to the channel data sets at another set of antenna arrays at a totally different location and a totally different frequency band was performed in \cite{R37R37}. According to this type of novel channel-to-channel mapping the authors efficiently guaranteed remarkable gains for MaMIMO systems. In particular, in
FDD MaMIMO, the UL channel data sets can be theoretically mapped to the DL ones. This is also fine in addition to a mapping of the DL channel coefficients at a given subset
of antenna arrays to the DL ones at all the other antenna sets. According to this, a considrable-and-efficient relaxation to the DL training/feedback overhead was totally proven to be guaranteed. For cell-free/distributed MaMIMO systems, furthermore, this novel kind of channel mapping is interpreted as follows. This is an efficient alleviation to the front-haul signaling overhead and even an extreme range of the generalised scenarios and problem-and-solutions.

\begin{figure}[t]
\centering
{\includegraphics[trim={{62mm} {100 mm} {52 mm} {100mm}},clip,scale=0.5]{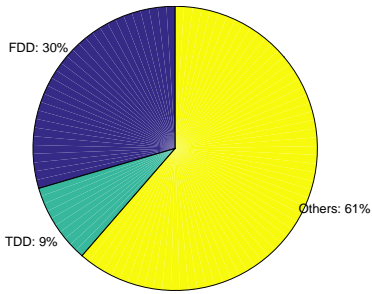}} 
\caption{FDD vs. TDD in DLg-based MaMIMO channel modeling/estimation.}  \label{F7}
\label{fig:EcUND} 
\end{figure} 
\begin{figure}[t]
\centering
{\includegraphics[trim={{62mm} {90 mm} {52 mm} {87mm}},clip,scale=0.5]{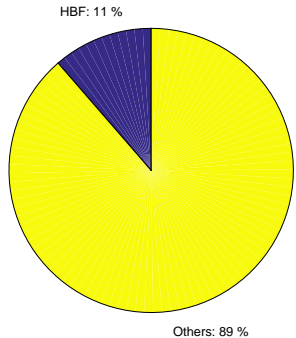}} 
\caption{HBF in DLg-based MaMIMO channel modeling/estimation.}  \label{F8}
\label{fig:EcUND} 
\end{figure} 
\begin{figure}[t]
\centering
{\includegraphics[trim={{57mm} {100 mm} {52 mm} {90mm}},clip,scale=0.5]{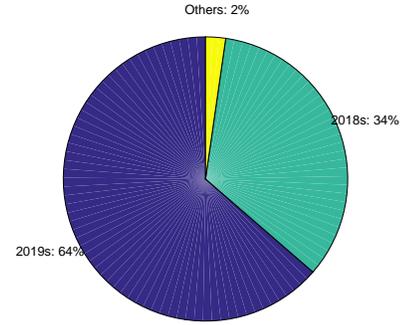}} 
\caption{DLg-based MaMIMO over the time.}  \label{F9}
\label{fig:EcUND} 
\end{figure} 

A DNN was technically proposed in \cite{R38R38} by initially demising the received signal, followed by a conventional least-squares estimation. This scheme efficiently works the same as MMSE estimator for high-dimensional signals, inversely, MMSE's requirement for complex channel inversions as well as the statistical knowledge of the CSI are not mandatory. The proposed method while it was efficiently proven that it is not necessary any training, this has also a robustness to pilot-contamination. This is also valid even if the number of antenna sets, subcarriers and coherence time interval (in terms of OFDM symbols) increase under a synchronisation error at the BS. The proposed method is really robust in the sense that it can perfectly remove the interference even if the eigen-space of the desired user and interference are strongly correlated. 

Offline-learning and online-learning techniques were applied to efficiently train the statistics of the wireless channel and the spatial structures in angle domain in \cite{R39R39}.

In \cite{R40R40}, a channel estimation DLg-based framework was essentially realised via outdated pilot. A DNN was efficiently exploited for channel prediction, where HBF was theoretically characterised. For frequency selective interference channels, the signal dimension is basically guaranteed, where a relaxation was theoretically characterised in relation to the interference. In addition, the analog RF beamformer
coefficients are fed to the time-domain signal.

Where the number of transmit antenna sets is greater than that of receive antenna ones, a new scheme was essentially realised in \cite{R41R41}. The projected gradient descent
method with trainable parameters was done, named the trainable projected gradient-detector. This efficient detector contains two iterative steps: the gradient descent phase and the soft projection one.

In \cite{R42R42}, in order to efficiently estimate the power allocation profiles
for a set of UEs' positions, how to learn a map between the positions of UEs and the optimal power allocation policies is essentially discovered.

In \cite{R43R43}, how to fully prove that the user-cell association is applicable in a real-time manner in practice in MaMIMO was efficiently explored, by using a DLg framework. 
The optimal user-cell association relying only upon the mobile users' positions
was theoretically proposed.

\begin{table}
\captionof{table}{{TDD vs. FDD in DLg based MaMIMO detection: main focus.}}
 \label{table2} 
\begin{tabular}{|p{4.5cm}|p{2.15cm}| }
 \hline    
 \hline    
\textsc{\textbf{TDD}} &   \textsc{\textbf{FDD}}  \\
 \hline  
  Pilot Contamination  &  CS Failure   \\
 \hline  
\end{tabular}
\end{table}
\begin{table}
\captionof{table}{{TDD vs. FDD in DLg based MaMIMO detection.}}
 \label{table3} 
\begin{tabular}{|p{2.5cm}|p{5.15cm}| }
 \hline    
 \hline    
\textsc{\textbf{TDD}} &   \textsc{\textbf{FDD}}  \\
 \hline  
  \cite{R21R21},  \cite{R22R22},   \cite{R23R23},   \cite{R24R24}  &  \cite{R1R1, R2R2, R4R4, R5R5, R6R6, R7R7, R8R8, R9R9, R10R10, R11R11, R12R12, R13R13, R14R14}   \\
 \hline  
\end{tabular}
\end{table}

A DLg approach was theoretically proposed \cite{R44R44} to efficiently predicte the UL channels for mixed analog-to-digital converters (ADCs), while a portion of antennas exploit low-resolution ADCs at the BS. The remaining antenna sets are essentially equipped with high-resolution ones. Only the signals received by the high-resolution ADC antenna sets are fully applied to efficiently estimate the channels of other antenna arrays in addition to their own channels. The out-performance of the proposed method over the existing ones, particularly mixed one-bit ADCs are completely shown. In MaMIMO, aimed at efficiently guaranteeing the hardware-cost reduction, low-resolution ADCs e.g. $1$-to-$3$ bits, are emerging. However, the low-resolution ADCs in practice experience the severely nonlinear distortion, in data detection. The mixed-ADCs are, instead, the candidates to efficiently reduce the hardware complexity, inversely, their overall-throughput, particularly for mixed one-bit ADC still lacks. A DLg based channel estimation mechanism was technically proposed in \cite{R44R44} for mixed-ADC MaMIMO UL. A novel estimation mapping from the channels of high-resolution ADC antenna sets to those of low-resolution ADC antenna arrays was efficiently proposed. This was theoretically characterised aimed at providing a relaxation to the detrimental knock-on effects of the severely distorted signals quantised by the low-resolution ADCs on the prediction accuracy. This framework is also implementable for the case with fewer high-resolution ADC antenna sets or correspondingly low signal-to-noise ratio (SNR).

In \cite{R45R45} a novel DLg based configuration was technically proposed by considering the correlated MaMIMO channel as a 3-D image. 

How to enhance BW for cell-free MaMIMO transceivers in mmWave bands which requires a high computational complexity was discussed in \cite{R46R46} in-depth. Towards this end, an accurate CSI prediction method was technically proposed. Indeed, a CNN based scheme was essentially realised totally suitable for an enormous number of SNR levels as the relative input. 

\section{Numerical results}
\subsection{Evaluation}
Although there seems to be no difference between TDD and FDD from a system-level-simulation point-of-view, as depicted in Fig. \ref{F6} \footnote{Fig. \ref{F6} is provided by LTE-Sim \cite{R47R47} where the number of cells is $1$ with the radius of $1 \; km$; the maximum-delay is $0.1$; and the users' speed is $3$ as pedestrians; while using $2$ users; doing the simulation for one time; for the widely accepted scheduling strategy proportional fairness which is called in the figure as \textit{PF}. In addition, the default scenario of \textit{Single-Cell-With-Interference} in LTE-Sim is used.}, there are some main challenges expressed in Table \ref{table2} for MaMIMO channel detection. 
In Fig. \ref{F7}, a comparison between FDD and TDD in MaMIMO transceivers is made. It is illustrated that FDD has been considered over than three times higher than that of the TDD technique. In addition, Fig. \ref{F9} also shows the DLg in MaMIMO channel modeling/estimation over the time. Table \ref{table3} also shows the work in which TDD and FDD were taken into consideration.

Fig. \ref{F8} shows HBF in DLg based MaMIMO channel modeling/prediction.

Fig. \ref{F9} shows DLg in MaMIMO systems over the time. It is revealed that recently, more specifically, in the time interval of 2018-2019, DLg techniques in MaMIMO channel modeling/estimation have given more attention.

Table \ref{table4} compares SLg with ULg for MaMIMO channel detection (although RLg was done in \cite{R28R28}).

\begin{table}
\captionof{table}{{SLg vs. ULg in MaMIMO channel detection.}}
 \label{table4} 
\begin{tabular}{|p{4.5cm}|p{2.15cm}| }
 \hline    
 \hline    
\textsc{\textbf{SLg}} &   \textsc{\textbf{ULg}}  \\
 \hline  
  \cite{R23R23, R36R36}  &   \cite{R35R35}  \\
 \hline  
\end{tabular}
\end{table}

\subsection{Open trends (Future work)}
\subsubsection{Q-learning}
After carefully reviewing the literature, it is revealed that no work has been done so far over Q-learning. A possible framework is then an adversarial-based one\footnote{See e.g. \cite{ R48R48, R49R49} to understand adversarial attacks to MaMIMO.} in which the attacker decided to be a \textit{Byzantine} node lying his channel-quality-indicator\footnote{LTE and 3gpp frequently consider this digit among the interval of [0-15], its maximum is conversely higher for 5G.} as a higher digit. This kind of perturbing is supposed to be performed aimed at guaranteeing a higher order of modulation, a higher coding-rate, and a considerably more acceptable block-error-rate. Towards this end, it is also important to note to the principle of frequency-synchronisation which provides a trade-off in relation to the falsification described above. 

\textit{What if we decide to provide a satisfactory solution to the data-pilot trade-off from a game-theoretical point of view? } Let us define a two-stage \textit{Stackelberg} game for two groups of agents: (\textit{i}) the data vector; and (\textit{ii}) the pilot vector. There arises a clearly momentous question how to enable players to estimate their rewards - the optimal value of which is the \textit{Value function} - using DLg. Let the utility function $\mathcal{U}$ be
\begin{equation*}
\mathcal{U}=\mathop{{\rm arg \; max}}\limits_{\mbox{\boldmath${d}$},\mbox{\boldmath${p}$}} {\rm \; }\left \{ \left \{\mathcal{R}^{(t)}_d (\cdots;\mbox{\boldmath${d}$})-\mathcal{R}^{(t)}_p (\cdots;\mbox{\boldmath${p}$}) \right \} , \left \{\mathcal{R}^{(t)}_p (\cdots;\mbox{\boldmath${p}$})-\mathcal{R}^{(t)}_d (\cdots;\mbox{\boldmath${d}$}) \right \} \right \},
\end{equation*}
according to which the data vector $\mbox{\boldmath${d}$}$ and the pilot vector $\mbox{\boldmath${p}$}$ are found. Meanwhile, for the $t$-th step, $\mathcal{R}^{(t)}_d (\cdots;\mbox{\boldmath${d}$})$ is the reward for $\mbox{\boldmath${d}$}$ and $\mathcal{R}^{(t)}_p (\cdots;\mbox{\boldmath${p}$})$ is the reward for $\mbox{\boldmath${p}$}$ which are the functions of some parameters as well as respectively $\mbox{\boldmath${d}$}$ and $\mbox{\boldmath${p}$}$. Let the aforementioned rewards be the rate.

\textit{\textbf{Fact 1 \cite{R51R51, R52R52}:}} Consider $\mathcal{S}$, $\mathcal{A}_z$ and $\mathcal{U}_z$ respectively as a finite set of Players, a set of Actions of the $z$-th Player, and the Utility Function for the $z$-th Player. A Game $\left \{ \mathcal{S}, \mathcal{A}_z,  \mathcal{U}_z \right \}$ has a \textit{Stackelberg Equilibrium} if: (\textit{i}) $\mathcal{A}_z, \forall z \in Z$ is a non-empty compact convex subset over the Euclidian Space; and (\textit{ii}) $\mathcal{U}_z$ is quasi-concave over $\mathcal{A}_z$. Additionally, the vector of Action Space $\mathcal{A}=( \mathbf{\tilde{a}}_{1}, \mathbf{\tilde{a}}_{2})=(\tilde{a}_{11}, \tilde{a}_{12}, \tilde{a}_{21}, \tilde{a}_{22})$ is a Stackelberg Equilibrium if and only if:
$\mathbf{\tilde{a}}_{1}=arg \;\mathop{{ max}} \limits_{\textbf{a}_{1}} \mathcal{U}_1 (\textbf{a}_{1}, \hat{\textbf{a}}_{2}(\textbf{a}_{1}))$ holds, where $\hat{\textbf{a}}_{2}(\textbf{a}_{1})=arg \;\mathop{{ max}} \limits_{\textbf{a}_{2}} (\textbf{a}_{1},\textbf{a}_{2}), \forall \textbf{a}_{1}$, and $\mathbf{\tilde{a}}_{2}=\hat{\textbf{a}}_{2}(\mathbf{\tilde{a}}_{1}).$ $\; \; \; \;   \square$

\textit{\textbf{ Claim 1:}} Our Stackelberg game is a \textit{two-stage} one since there seems to be two separately defined controllers, that is, two rewards given above. These controllers are assigned to two separate group-Players, i.e., the drays of $\mbox{\boldmath${d}$}$ and $\mbox{\boldmath${p}$}$. $\; \; \; \;   \square$

Now we need to re-write reward as a $\mathcal{Q}$-function in order to map it. Thus, the $\mathcal{Q}$-functions in relation to $\mbox{\boldmath${d}$}$ and $\mbox{\boldmath${p}$}$ are respectively $\mathcal{Q}_d(e^{(d)},a^{(d)}_z)=w^{(d)T}_z\mbox{\boldmath${x}$}^{(d)}+b^{(d)}_z, w^{(d)}_z(t+1)=w^{(d)}_z(t)+\lambda (\mathcal{R}_d^{(*)}-\mathcal{Q}_d(e^{(d)},a^{(d)}_z))$ and $\mathcal{Q}_p(e^{(p)},a^{(p)}_z)=w^{(p)T}_z\mbox{\boldmath${x}$}^{(p)}+b^{(p)}_z, w^{(p)}_z(t+1)=w^{(p)}_z(t)+\lambda (\mathcal{R}_p^{(*)}-\mathcal{Q}_p(e^{(p)},a^{(p)}_z))$ while $(\cdot)^T$ stands for the transpose operand and $\lambda$ is the leaning rate, w.r.t. the optimal value functions $\mathcal{R}_d^{(*)}$ and $\mathcal{R}_p^{(*)}$. Towards this end, we need to update the weights $w^{(d)}_z$ and $w^{(p)}_z$, as well as the biases $b^{(d)}_z$ and $b^{(p)}_z$. Meanwhile, $e^{(\cdot)}$ is the partial state observed by the $z$-th agent and $a^{(\cdot)}_z$ is the action relating to her.
Finally, $a^{(\cdot)}_z(e^{(\cdot)})=\mathop{{\rm arg \; max}}\limits_{z} \left \{ \mathcal{Q}_{\cdot}(e^{(\cdot)},a^{(\cdot)}_z) \right \}$ emphatically provides a strongly affirmative response to the open trend discussed in this part, i.e., Q-learning.

\begin{figure}[t]
\centering
{\includegraphics[trim={{40mm} {86 mm} {42 mm} {90mm}},clip,scale=0.5]{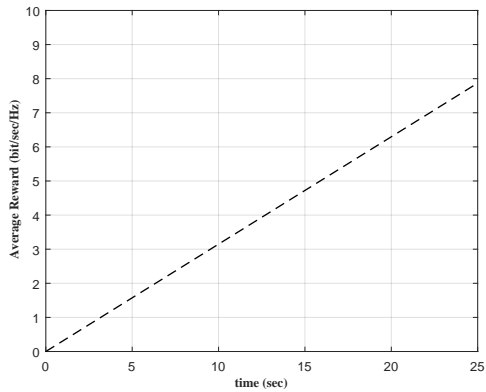}} 
\caption{DLg-based MaMIMO over the time.}  \label{F10}
\label{fig:EcUND} 
\end{figure}

Fig. \ref{F10} shows the average reward (rate for the data set) versus time for a simple scenario. 

\subsubsection{Frequency synchronisation}
As discussed in \cite{R50R50}, one entirely central issue is the principle of \textit{frequency synchronisation}. This is totally important while doing the carrier-frequency-offset estimation for the MaMIMO system. Therefore, one possible work is to define a new DLg based framework in this context.

\subsubsection{Pilot contamination}
In \cite{R22R22}, the issue of pilot contamination was totally solved under the assumption of a perfect knowledge of the adjacent-cells' pilots. One may physically examine an imperfect knowledge, e.g. w.r.t. an adversarial attack, in this context.

\subsubsection{$L_s$ vs $N_t$}
Another possible work in the future is some information-theoretic solutions to the tradeoff between the two controversial conditions $ L_s < N_t$ and $ L_s \nless N_t$, as discussed in \cite{R13R13}.

\subsubsection{complexity vs. accuracy}
\textsc{\textbf{Remark 3-}} \textit{Among tens of references in the literature, the sole references in which both accuracy and complexity were analysed in parallel with each other (even compared to other work) are listed as follows: \cite{R6R6, R12R12, R13R13, R17R17, R18R18, R19R19, R21R21, R22R22, R23R23, R33R33, R35R35, R38R38, R41R41, R42R42, R43R43}.}

\textit{Remark 3} shows a possible field in the context of new architectures and algorithms which are both practical and accurate. 

\subsubsection{1-ADC and ADC distortions}
One possible work is also taking into account the ADC impairments and more realistic and practical bottlenecks for the MaMIMO transceiver.

\subsubsection{reciprocity between UL and DL in practice}
As discussed in the previous parts (\cite{R4R4}), the reciprocity between UL and DL hardly does it hold. Inversely, the major part of the literature as discussed in the previous parts (even \cite{R53R53, R54R54}), do not focus on this point, therefore, this may be an open field if we focus on it. 

\section{conclusion}
A comprehensive survey over those challenges in MaMIMO channel modeling that were solved via DLg techniques was performed. The main focus of the recently published papers was on the outperformance of DLg methods over CS ones. This was because, as discussed, there seems to be five crucial problems in relation to CS techniques in MaMIMO channel modeling and estimation. The aforementioned problems, specifically, the feedback-overhead, enforced the scholars and researchers in both Academia and Industry to pursue their trends towards DLg oriented mechanisms. Additionally, FDD has more challenging bottlenecks compared to TDD. More interestingly, we proposed a possible game-theoretical solution in terms of Q-learning – something that was for the first time.

%
\markboth{IEEE, VOL. XX, NO. XX, X 2019}%
{Shell \MakeLowercase{\textit{et al.}}: Bare Demo of IEEEtran.cls for Computer Society Journals}

\begin{thebibliography}{}
\bibitem{R1R1}Z. Jiang, S. Chen, A. F. Molisch, R. Vannithamby, "Exploiting Wireless Channel State Information Structures Beyond Linear Correlations: A Deep Learning Approach," \textit{IEEE Commun. Magazine}, Vol. 57, no. 3, pp. 28-34, 2019.
\bibitem{R2R2}C. Luo, J. Ji, Q. Wang, X. Chen, P. Li, "Channel State Information Prediction for 5G Wireless Communications: A Deep Learning Approach," \textit{IEEE Trans. Network Science. Eng.}, Vol. PP, no. 99, pp. 1-1, 2019.
\bibitem{R3R3}N. Tishby, F. C. Pereira, W. Bialek, "The information bottleneck method," https://arxiv.org/abs/physics/0004057, 2000.
\bibitem{R4R4}H. Sun, Z. Zhao, X. Fu, M. Hong, "Limited Feedback Double Directional Massive MIMO Channel Estimation: From Low-Rank Modeling to Deep Learning,"
\textit{In Proc. 2018 IEEE 19th I. Workshop Signal Proc. Advances Wireless Commun. (SPAWC)}, 25-28 June 2018, Kalamata, Greece.
\bibitem{R5R5}Z. Liu, L. Zhang, Z. Ding, "Exploiting Bi-Directional Channel Reciprocity in Deep Learning for Low Rate Massive MIMO CSI Feedback," \textit{IEEE Wireless Commun. Letters}, Vol. 8, no. 3, pp. 889-892, 2019.
\bibitem{R6R6}Y. Yang, F. Gao, G. Y. Li, M. Jian, "Deep Learning based Downlink Channel Prediction for FDD Massive MIMO System," \textit{IEEE Commun. Letters}, Vol. PP, no. 99, pp. 1-1, 2019.
\bibitem{R7R7}C. Qing, B. Cai, Q. Yang, J. Wang, C. Huang, "Deep Learning for CSI Feedback Based on Superimposed Coding," \textit{IEEE Access}, Vol. 7, pp. 93723-93733, 2019.
\bibitem{R8R8}T. Wang, C. Wen, S. Jin, G. Y. Li, "Deep Learning-Based CSI Feedback Approach for Time-Varying Massive MIMO Channels," \textit{IEEE Wireless Commun. Letters}, Vol. 8, no. 2, pp. 416-419, 2019.
\bibitem{R9R9}Z. Jiang, Z. He, S. Chen, A. F. Molisch, S. Zhou, "Inferring Remote Channel State Information: Cramer-Rae Lower Bound and Deep Learning Implementation," \textit{In Proc. IEEE Global Commun. Conf. (GLOBECOM)}, 9-13 Dec. 2018, Abu Dhabi, UAE.
\bibitem{R10R10}J. Wang, Y. Ding, S. Bian, Y. Peng, M. Liu, G. Gui, "UL-CSI Data Driven Deep Learning for Predicting DL-CSI in Cellular FDD Systems," \textit{IEEE Access}, Vol. 7, pp. 96105-96112, 2019.
\bibitem{R11R11}C. Wen, W. Shih, S. Jin, "Deep Learning for Massive MIMO CSI Feedback," \textit{IEEE Wireless Commun. Letters}, Vol. 7, no. 5, pp. 748-751, 2018.
\bibitem{R12R12}P. Wu, Z. Liu, J. Cheng, "Compressed CSI Feedback With Learned Measurement Matrix for mmWave Massive MIMO," https://arxiv.org/abs/1903.02127, 2019.
\bibitem{R13R13}C. Chun, J. Kang, I. Kim, "Deep Learning-Based Channel Estimation for Massive MIMO Systems," \textit{IEEE Wireless Commun. Letters}, Vol. 8, no. 4, pp. 1228-1231, 2019.
\bibitem{R14R14}G. Gao, C. Dong, K. Niu, "Sparsely Connected Neural Network for Massive MIMO Detection," \textit{In Proc. 2018 IEEE 4th I. Conf. Computer and Commun. (ICCC)}, 7-10 Dec. 2018, China.
\bibitem{R15R15}N. Samuel, T. Diskin and A. Wiesel, "Deep MIMO detection," \textit{In Proc. 2017 IEEE 18th I. W. Signal Proc. A. Wireless Commun. (SPAWC)}, Sapporo, 2017.
\bibitem{R16R16}H. He, C. Wen, S. Jin, G. Y. Li, "Deep Learning-Based Channel Estimation for Beamspace mmWave Massive MIMO Systems," \textit{IEEE Wireless Commun. Letters}, Vol. 7, no. 5, pp. 852-855, 2018.
\bibitem{R17R17}P. Dong, H. Zhang, G. Y. Li, N. NaderiAlizadeh, and I. S. Gaspar, "Deep CNN based Channel Estimation for mmWave Massive MIMO Systems," \textit{in Proc. IEEE I. Conf. Acoustics, Speech and Signal Proc. (ICASSP)}, Brighton, UK, 12-17 May 2019.
\bibitem{R18R18}D. Hu, Y. Zhang, L. He, J. Wu, "Low-Complexity Deep-Learning-Based DOA Estimation for Hybrid Massive MIMO Systems with Uniform Circular Arrays," \textit{IEEE Wireless Commun. Letters}, Vol. PP, no. 99, pp. 1-1, 2019.
\bibitem{R19R19}S. Gao, X. Cheng, L. Yang, "Making Wideband Channel Estimation Feasible for mmWave Massive MIMO: A Doubly Sparse Approach,"
\textit{In Proc. 2019 IEEE I. Conf. Commun. (ICC)}, 20-24 May 2019, China.
\bibitem{R20R20}J. Joo, M. Chul Park, D. S. Han, V. Pejovic, "Deep Learning-Based Channel Prediction in Realistic Vehicular Communications," \textit{IEEE Access}, Vol. 7, pp. 27846-27858, 2019.
\bibitem{R21R21}H. Huang, G. Gui, H. Sari, F. Adachi, "Deep Learning for Super-Resolution DOA Estimation in Massive MIMO Systems,"
\textit{In Proc. 2018 IEEE 88th Vehicular Techno. Conf. (VTC-Fall)}, 27-30 Aug. 2018, IL, USA.
\bibitem{R22R22}M. Borgerding, P. Schniter, S. Rangan, "AMP-Inspired Deep Networks for Sparse Linear Inverse Problems," \textit{IEEE Trans. Signal Proc.}, Vol. 65, no. 16, pp. 4293-4308, 2017.
\bibitem{R23R23}K. Kim, J. Lee, J. Choi, "Deep Learning Based Pilot Allocation Scheme (DL-PAS) for 5G Massive MIMO System," \textit{IEEE Commun. Letters}, Vol. 22, no. 4, pp. 828-831, 2018.
\bibitem{R24R24}C. Huang, G. C. Alexandropoulos, A. Zappone, C. Yu, "Deep Learning for UL/DL Channel Calibration in Generic Massive MIMO Systems," \textit{In Proc. 2019 IEEE I. Conf. Commun. (ICC)}, 20-24 May 2019, China.
\bibitem{R25R25}Y. Liao, H. Yao, Y. Hua, C. Li, "CSI Feedback Based on Deep Learning for Massive MIMO Systems," \textit{IEEE Access}, Vol. 7, pp. 86810-86820, 2019.
\bibitem{R26R26}Y. Zhou, Z. Md. Fadlullah, B. Mao, N. Kato, "A Deep-Learning-Based Radio Resource Assignment Technique for 5G Ultra Dense Networks," \textit{IEEE Network}, Vol. 32, no. 6, pp. 28-34, 2018.
\bibitem{R27R27}Y. Yang, Y. Li, K. Li, S. Zhao, R. Chen, J. Wang, "DECCO: Deep-Learning Enabled Coverage and Capacity Optimization for Massive MIMO Systems," \textit{IEEE Access}, Vol. 6, pp. 23361-23371, 2018.
\bibitem{R28R28}T. Maksymyuk, J. Gazda, O. Yaremko, D. Nevinskiy, "Deep Learning Based Massive MIMO Beamforming for 5G Mobile Network,"
\textit{In Proc. 2018 IEEE 4th I. S. Wireless Systems I. Conf. Intelligent Data Acquisition Advanced Computing Systems (IDAACS-SWS)}, 20-21 Sept. 2018, Lviv, Ukraine.
\bibitem{R29R29}X. Li, X. Yu, T. Sun, J. Guo, J. Zhang, "Joint Scheduling and Deep Learning-Based Beamforming for FD-MIMO Systems Over Correlated Rician Fading," \textit{IEEE Access}, Vol. 7, pp. 118297-118309, 2019.
\bibitem{R30R30}M. Arnold, J. Hoydis, S. Brink, "Novel Massive MIMO Channel Sounding Data applied to Deep Learning-based Indoor Positioning,"
\textit{In Proc. 12th International ITG Conf. Systems, Commun. Coding}, 11-14 Feb. 2019, Rostock, Germany.
\bibitem{R31R31}M. Arnold, S. Dorner, S. Cammerer, S. Ten, "On Deep Learning-Based Massive MIMO Indoor User Localization," \textit{In Proc. 2018 IEEE 19th I. W. Signal Proc. Advances Wireless Commun. (SPAWC)}, 25-28 June 2018, Kalamata, Greece.
\bibitem{R32R32}S. Chen, Z. Jiang, S. Zhou, Z. Niu, "Learning-Based Remote Channel Inference: Feasibility Analysis and Case Study," \textit{IEEE Trans. Wireless Commun.}, Vol. 18, no. 7, pp. 3554-3568, 2019.
\bibitem{R33R33}H. Huang, Y. Song, J. Yang, G. Gui, F. Adachi, "Deep-Learning-Based Millimeter-Wave Massive MIMO for Hybrid Precoding," \textit{IEEE Trans. Vehicular Technol.}, Vol. 68, no. 3, pp. 3027-3032, 2019. 
\bibitem{R34R34}S. Takabe, M. Imanishi, T. Wadayama, K. Hayashi, "Deep Learning-Aided Projected Gradient Detector for Massive Overloaded MIMO Channels,"
\textit{In Proc. 2019 IEEE I. Conf. Commun. (ICC)}, 20-24 May 2019, China.
\bibitem{R35R35}J. Xu, P. Zhu, J. Li, X. You, "Deep Learning-Based Pilot Design for Multi-User Distributed Massive MIMO Systems," \textit{IEEE Wireless Commun. Letters}, Vol. 8, no. 4, pp. 1016-1019, 2019.
\bibitem{R36R36}T. V. Chien, E. Bjornson, E. G. Larsson, "Sum Spectral Efficiency Maximization in Massive MIMO Systems: Benefits from Deep Learning," \textit{In Proc. 2019 IEEE I. Conf. Commun. (ICC)}, 20-24 May 2019, China.
\bibitem{R37R37}M. Alrabeiah, A. Alkhateeb, "Deep Learning for TDD and FDD Massive MIMO: Mapping Channels in Space and Frequency," https://arxiv.org/abs/1905.03761, 2019.
\bibitem{R38R38}E. Balevi, A. Doshi, J. G. Andrews, "Massive MIMO Channel Estimation with an Untrained Deep Neural Network," https://arxiv.org/abs/1908.00144, 2019.
\bibitem{R39R39}H. Huang, J. Yang, H. Huang, Y. Song, G. Gui, "Deep Learning for Super-Resolution Channel Estimation and DOA Estimation Based Massive MIMO System," \textit{IEEE Trans. Vehicular Technol.}, Vol. 67, no. 9, pp. 8549-8560, 2018. 
\bibitem{R40R40}K. Satyanarayana, M. El-Hajjar, A. A. M. Mourad, L. Hanzo, "Multi-User Full Duplex Transceiver Design for mmWave Systems Using Learning-Aided Channel Prediction," \textit{IEEE Access}, Vol. 7, pp. 66068-66083, 2019.
\bibitem{R41R41}S. Takabe, M. Imanishi, T. Wadayama, R. Hayakaw, "Trainable Projected Gradient Detector for Massive Overloaded MIMO Channels: Data-Driven Tuning Approach," \textit{IEEE Access}, Vol. 7, pp. 93326-93338, 2019.
\bibitem{R42R42}L. Sanguinetti, A. Zappone, M. Debbah, "Deep Learning Power Allocation in Massive MIMO," \textit{In Proc. 2018 52nd Asilomar Conf. Signals, Systems, and Computers}, 28-31 Oct. 2018, CA, USA.
\bibitem{R43R43}A. Zappone, L. Sanguinetti, M. Debbah, "User Association and Load Balancing for Massive MIMO through Deep Learning,"
in: 2018 52nd Asilomar Conference on Signals, Systems, and Computers
Date of Conference: 28-31 Oct. 2018 rence Location: Pacific Grove, CA, USA, USA
\bibitem{R44R44}S. Gao, P. Dong, Z. Pan, G. Y. Li, "Deep Learning based Channel Estimation for Massive MIMO with Mixed-Resolution ADCs," \textit{IEEE Commun. Letters}, Vol. PP, no. 99, pp. 1-1, 2019.

\bibitem{R45R45}J. I. Chen, K. L. Lai, "Reduce the Correlation Phenomena over Massive-MIMO System by Deep Learning Algorithms," \textit{In Proc. 2018 IEEE I. Conf. Advanced Manufacturing (ICAM)}, 16-18 Nov. 2018.
\bibitem{R46R46}Y. Jin, J. Zhang, S. Jin, B. Ai, "Channel Estimation for Cell-Free mmWave Massive MIMO Through Deep Learning," \textit{IEEE Trans. Vehicular Technol.}, Vol. PP, no. 99, pp. 1-1, 2019.
\bibitem{R47R47}G. Piro, L. A. Grieco, G. Boggia, F. Capozzi, and P. Camarda, "Simulating
LTE cellular systems: an open-source framework," \textit{IEEE Trans. Vehicular Technol.}, vol. 60, no. 2, pp. 498-513, 2011.
\bibitem{R48R48}M. Sadeghi, E. G. Larsson, "Physical Adversarial Attacks Against End-to-End Autoencoder Communication Systems," \textit{IEEE Commun. Letters}, Vol. 23, no. 5, pp. 847-850, 2019.
\bibitem{R49R49}M. Sadeghi, E. G. Larsson, "Adversarial Attacks on Deep-Learning Based Radio Signal Classification," \textit{IEEE Wireless Commun. Letters}, Vol. 8, no. 1, pp. 213-216, 2019.
\bibitem{R50R50}W. Zhang, F. Gao, S. Jin, H. Lin, "Frequency Synchronization for Uplink Massive MIMO Systems," \textit{IEEE Trans. Wireless Commun.} Vol. 17, no. 1, pp. 235-249, 2018.

\bibitem{R51R51}Osborne; M. J.; Rubenstein, A. A course in game theory. \textit{Cambridge: MIT Press.}, 1994.
\bibitem{R52R52}M. Haddad; Y. Hayely; and O. Habachi. Spectrum Coordination in Energy Efficient Cognitive Radio Networks. \textit{IEEE Trans. Veh. Technol.}, Vol. 64, pp. 2112-2122, 2015.

\bibitem{R53R53}MB. Mashhadi, Q. Yang, D. Gunduz, "CNN-based Analog CSI Feedback in FDD MIMO-OFDM Systems,"
https://arxiv.org/abs/1910.10428, 2019.
\bibitem{R54R54}Q. Yang, MB. Mashhadi, D. Gunduz, "Deep Convolutional Compression for Massive MIMO CSI Feedback," https://arxiv.org/abs/1907.02942, 2019.




\end{thebibliography}
\end{document}